\documentclass{emulateapj}

\newcommand{\rxjw}{RX~J1856.5$-$3754}
\newcommand{\rxjk}{RX~J0720.4$-$3125}
\newcommand{\RXJ}{RX~J2143.0+0654}
\newcommand{\rbs}{RX~J1308.6+2127}
\newcommand{\rxj}{RX~J2143}

\newcommand{\xmm}{{\em XMM}}
\defcitealias{kvk05}{KvK05a}
\defcitealias{kvk05b}{KvK05b}
\defcitealias{vkk08}{vKK08}
\defcitealias{zct+05}{Z+05}
\newcommand{\Hzsec}{\ensuremath{{\rm Hz}\,{\rm s}^{-1}}}
\newcommand{\secsec}{\ensuremath{{\rm s}\,{\rm s}^{-1}}}
\newcommand{\expnt}[2]{\ensuremath{#1 \times 10^{#2}}}   

\begin{document}

\shorttitle{The Spin-down of the Nearby Isolated Neutron Star \RXJ}
\shortauthors{Kaplan \& van~Kerkwijk}

\title{Constraining the Spin-down of the Nearby Isolated Neutron Star \RXJ\altaffilmark{1}}

\author{D.~L.~Kaplan\altaffilmark{1} and M.~H.~van  Kerkwijk\altaffilmark{2}}
\altaffiltext{1}{Hubble Fellow; KITP, Kohn Hall, University of
  California, Santa Barbara, CA 93106; dkaplan@kitp.ucsb.edu}

\altaffiltext{2}{Department of Astronomy and Astrophysics, University
  of Toronto, 50 St.\ George Street, Toronto, ON M5S 3H4, Canada;
  mhvk@astro.utoronto.ca}

\slugcomment{Accepted for publication in ApJ Letters}
\begin{abstract}
  Magnetic field estimates for nearby isolated neutron stars (INS) help to
  constrain both the characteristics of the population and the nature
  of their peculiar X-ray spectra.  From a series of
  \xmm\textit{-Newton}\ observations of \RXJ, we measure a spin-down
  rate of $\dot\nu = \expnt{(-4.6\pm2.0)}{-16}\,\Hzsec$.  While this
  does not allow a definitive measurement of the dipole magnetic field
  strength, fields of $\gtrsim\!10^{14}\,$G such as those inferred from
  the presence of a spectral absorption feature at 0.75\,keV are
  excluded.  Instead, the field is most likely around
  $\expnt{2}{13}\,$G, very similar to those of other INS.  We not only
  suggest that this similarity most likely reflects the
  influence of magnetic field decay on this population, but also
  discuss a more speculative possibility that it results from peculiar
  conditions on the neutron-star surface.  We find no evidence for
  spectral variability above the $\sim\!2$\% level.  We 
  confirm the presence of the 0.75-keV feature found earlier, and
  find tentative evidence for an additional absorption feature at
  0.4\,keV.
\end{abstract}
\keywords{stars: individual (RX J2143.0+0654) ---
  stars: neutron --- X-rays: stars}

\section{Introduction}
The so-called isolated neutron stars (INS; see \citealt{haberl07} and
\citealt{kaplan08} for reviews) are a group of seven nearby
($\lesssim\!1\,$kpc) neutron stars with low
($\sim\!10^{32}{\rm\,erg\,s^{-1}}$) X-ray luminosities and long
(3--10\,s) spin periods.  When first discovered, a range of possible
reasons for their long periods was suggested, with a corresponding
range in magnetic field strengths of $10^{10}$--$10^{15}\,$G :
accretors \citep{kp97,w97}, long-period pulsars
\citep{kvk98,kkvkm02,zhc+02}, and middle-aged magnetars \citep{hk98}.
Field measurements can allow one to distinguish between these
possibilities, and thus help understand the INS and their place in the
greater neutron-star population.  The magnetic field is also a
necessary component of any realistic thermal emission models
\citep{ztd04,mzh03,hkc+07}, required to interpret the surface emission
and deduce radii and other properties of the INS.

Dipolar magnetic field strengths can be estimated from coherent timing
solutions, and we  used X-ray observations to derive such
solutions for three INS,\footnote{Our provisional solution for \rxjw\
\citepalias{vkk08} was confirmed by two additional observations taken
in 2008 March and October.} finding magnetic fields of
$\expnt{(1-3)}{13}\,$G 
(\citealt{kvk05,kvk05b}, hereafter \citetalias{kvk05},b;
\citealt{vkk08}, hereafter \citetalias{vkk08};  also see \citealt{vkkpm07}).
This breakthrough is complemented by the discovery of broad absorption
features at energies of 0.2--0.75\,keV in the thermal (with bolometric
luminosities of $\sim 100$ times the spin-down luminosity, the emission is
certainly thermal) spectra of six of the seven INS.  Assuming a pure
hydrogen atmosphere, fields of $10^{13}$--$10^{14}\,$G, and
temperatures around $10^6\,$K, the absorption could be due to either
proton cyclotron or transitions between bound states of neutral
hydrogen.  Intriguingly, for the two objects with both spectroscopic
and spin-down magnetic fields, the agreement appeared to be good
\citep{vkk07}.

Here, we constrain the spin-down rate and hence magnetic field
strength of the INS \object[RX J2143.0+0654]{\RXJ}\ (also
1RXS~J214303.7+065419 or RBS~1774; hereafter \rxj) with a series of
dedicated \xmm-\textit{Newton}\ observations.
\rxj\ was identified as a possible neutron
star by \citet{zct+01} on the basis of a  soft thermal spectrum
and the absence of an  optical counterpart.  Using \xmm,
\citet[][hereafter \citetalias{zct+05}]{zct+05} confirmed that the
spectrum was soft and blackbody-like.  They also found a broad
absorption feature around 0.75$\,$keV and identified a candidate
$9.437\,$s periodicity.  For hydrogen atmospheres, the high energy of
the absorption (considerably higher than those of the other INS; see
\citealt{haberl07} and \citealt{vkk07}) should translate to a high
magnetic field of $\gtrsim\!10^{14}\,$G, which makes this object a
particularly good test for models of the origin of the X-ray
absorption and for understanding the INS population.

\section{Observations and Analysis}
We observed \rxj\ 11 times with \xmm\ \citep{jla+01} in 2007 and
2008, and focus here on the data taken with the European Photon
Imaging Camera (EPIC) with pn and MOS detectors, all used in small
window mode with thin filters (Table~\ref{tab:obs}).  All our
observations, as well as the one from \citetalias{zct+05} (taken with
the same settings), were processed with SAS version 8.0.  We used {\tt
epchain} and {\tt emchain} and selected source events from a circular
region of $37\farcs5$ radius with energies below 2\,keV (where flares
are negligible; the source is not detected above 2\,keV).  As
recommended, we included only one- and two-pixel (single and double
patterns 0--4) events with no warning flags for pn, and single,
double, and triple events (patterns 0--12) with the default flag
mask for MOS1/2.
We barycentered the event times using the {\em Chandra X-ray
  Observatory} position from \citet{rtj+07}: $\alpha=21^{\rm h}43^{\rm
  m}03\fs38$ and $\delta=+06\degr54\arcmin17\farcs5$ (J2000).

\begin{deluxetable}{c c c c c}
\tablewidth{0pt}
\tablecaption{Log of Observations and Times of Arrival\label{tab:obs}}
\tablehead{
&&\colhead{Exp.\tablenotemark{a}}&&\colhead{TOA\tablenotemark{b}}\\
\colhead{Rev.}&
\colhead{Date}&\colhead{(ks)}&\colhead{Counts\tablenotemark{a}}& \colhead{(MJD)}\\[-2.2ex]
}
\startdata
\dataset[ADS/XMM\#0201150101]{\phn820} & 2004~May~31 & 30.0 & 50,032 & 53,156.821900(2)\\
\dataset[ADS/XMM\#0502040601]{1360} & 2007~May~13 & 12.9 & 22,915 & 54,233.762142(4)\\
\dataset[ADS/XMM\#0502040701]{1362} & 2007~May~17 & 13.0 & 22,926 & 54,237.950712(4)\\
\dataset[ADS/XMM\#0502040801]{1368} & 2007~May~30 & \phn7.0 & 12,369 & 54,250.632660(4)\\
\dataset[ADS/XMM\#0502040901]{1375} & 2007~Jun~12 & \phn8.3 & 14,497 & 54,263.916974(5)\\
\dataset[ADS/XMM\#0502041001]{1447} & 2007~Nov~03 & \phn8.5 & 15,049 & 54,407.453138(4)\\
\dataset[ADS/XMM\#0502041101]{1449} & 2007~Nov~07 & 11.1 & 19,480 & 54,411.246902(4)\\
\dataset[ADS/XMM\#0502041201]{1449} & 2007~Nov~08 & \phn9.5 & 16,687 & 54,412.213081(3)\\
\dataset[ADS/XMM\#0502041301]{1457} & 2007~Nov~23 & \phn5.3 & \phn9,229 & 54,427.463785(6)\\
\dataset[ADS/XMM\#0502041401]{1465} & 2007~Dec~10 & \phn7.7 & 13,517 & 54,444.074930(4)\\
\dataset[ADS/XMM\#0502041501]{1538} & 2008~May~03 & 11.5 & 20,812 & 54,589.198404(4)\\
\dataset[ADS/XMM\#0502041801]{1546} & 2008~May~19 & \phn7.7 & 13,414 & 54,605.222248(5)\\
\enddata
\tablecomments{All observations used EPIC-pn in the small window
  mode and with the thin filter for  both EPIC-pn and EPIC-MOS1/2.   }
\tablenotetext{a}{The exposure time and number of counts given here
  are for EPIC-pn only.}
\tablenotetext{b}{The TOA is defined as the time of maximum light of
  the pulsation closest to the middle of each observation computed
  from the combined EPIC-pn and EPIC-MOS1/2 datasets, and is given
  with 1-$\sigma$ uncertainties.}
\end{deluxetable}

\subsection{Timing Analysis}
We first determined the energy range that maximized the power in a
$Z_1^2$ periodogram for the combined EPIC data from 2004 (the longest
single observation).  We found that events below 130$\,$eV
($\sim\!23\%$ of all events) were only marginally pulsed, and that
removing those gave maximum power ($Z_1^2=58.2$; all events gives
56.9).  With these, we found a best-fit frequency of
$\nu=0.106066\pm0.000003\,$Hz
(our final timing solution does not change significantly
when we include the low-energy events).

\begin{figure}
\plotone{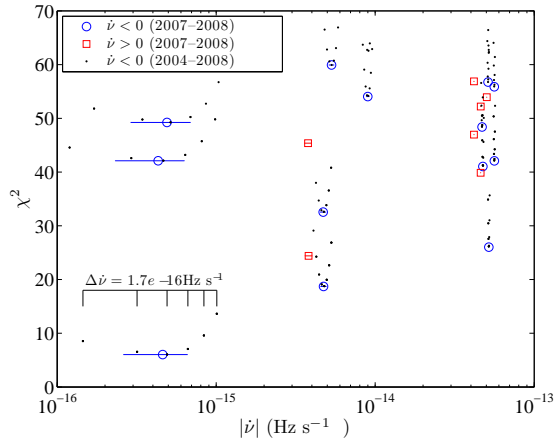}
\caption{Phase residuals for \rxj.  In the top panel, the residuals
  are shown relative to a linear model ($\dot\nu=0$).  The solid line
  and filled squares show the best-fit solution including the 2004
  data, while the dashed lines and open squares show the five alternate
  solutions that are within $2\sigma$ and have cycle counts for the
  2004 observation varying by 1 cycle (see text and
  Fig.~\ref{fig:solns}).  In the bottom panel, the residuals to the
  best-fit quadratic model are shown.}
\label{fig:resids}
\end{figure}

\begin{figure}
\plotone{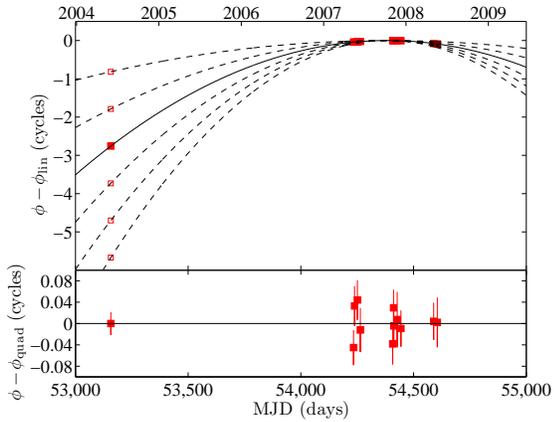}
\caption{Possible timing solutions for \rxj.  Shown are $\chi^2$
  values vs.\ spin-down rates $|\dot \nu|$ (along the bottom axis)
  and dipole magnetic field (top axis) with 1\,$\sigma$
  uncertainties for solutions with $\dot \nu<0$ (circles) and $\dot
  \nu>0$ (squares), using only the data from 2007--2008 (8 dof).  Also
  shown are the $\chi^2$ values for the $\dot \nu<0$ solutions
  incorporating the data from 2004 (9 dof; points), which have aliases
  with spacing $\expnt{1.7}{-16}\,\Hzsec$ around each of the primary
  solutions (as labeled).}
\label{fig:solns}
\end{figure}

Using the above frequency, we determined the times of arrival (TOAs;
see Table~\ref{tab:obs}) for the combined EPIC data from each
observation by fitting sine functions to the binned lightcurves
(following \citetalias{kvk05}).  The spacing and precision of the TOAs
is insufficient for an unambiguous timing solution (unlike in
\citetalias{kvk05},b but as in \citetalias{vkk08}).  Instead we
searched for possible coherent timing solutions by iteratively trying
sets of cycle counts between the different TOAs from 2007 and 2008
(similar to \citetalias{vkk08}, although we did not incorporate
frequency information here because it does not add extra information).
We find one solution (Fig.~\ref{fig:resids}) that is considerably
better than the alternatives (Fig.~\ref{fig:solns}) with $\chi^2=6.0$
for 8 degrees of freedom (dof) and a small, marginally significant
spin-down rate of $\dot \nu=\expnt{(-4.6\pm2.0)}{-16}\,\Hzsec$ for a
dipole magnetic field of $B_{\rm dip}\approx\expnt{2}{13}\,$G (see
Table~\ref{tab:ephem}).  The next best solution has $\chi^2=18.7$ and
a much larger spin-down rate: $\expnt{(-4.7\pm0.2)}{-15}\,\Hzsec$,
resulting from a difference of 1 cycle for the gap between Revs.~1465
and 1538.
For a fit with 3 free parameters, a change in $\chi^2$ of 12.7 means
that the best-fit solution is favored at 99.5\% confidence.  Even less
likely solutions are found for combinations of other cycle counts
(Fig.~\ref{fig:solns}).

We confirmed this solution in two ways.  First, we verified that an
analysis using the TOAs from the individual instruments gave the same
result.  Second, we did a coherent $Z_1^2$ periodogram as a function
of both $\nu$ and $\dot \nu$.  The best four solutions identified were
the same as those found in the TOA analysis.  The best-fit peak has
$Z_1^2=202.9$, consistent with the pulsed fraction of $\sim\!4$\%.
Spin-down is only marginally
detected at about 93\% confidence, consistent with the TOA analysis.

Unfortunately, we  cannot unambiguously extend the
solution back to the 2004 observation: given its offset $\Delta
t=1,233$\,days from the reference time $t_0$ of the above solution, the
uncertainty on the cycle count is approximately
$\frac{1}{2}\sigma_{\dot \nu}\Delta t^2=1.21\,$cycles.  As a result,
about six solutions are within $\pm 2\sigma$ of the best-fit solution,
with implied spin-down values differing by $\sigma_{\dot
  \nu}/1.21\,{\rm cycles}=\expnt{1.7}{-16}\,\Hzsec$
(Figs.~\ref{fig:resids} and \ref{fig:solns}).

\subsection{Spectral Analysis}
\label{sec:spec}
We examined the EPIC-pn spectra of \rxj\ from our new data to verify
whether the basic fits of \citetalias{zct+05} are still valid with our
$\sim\!2.5$ times longer total exposure time, and to look for possible
long-term variability such as what \citet{dvvmv04} found for \rxjk.
(A full spectral analysis,  including the EPIC-MOS and RGS data
and phase-resolved fits, is in progress.)  For our spectra, we
used the same extraction regions as for the timing analysis, and an
offset circular region with the same radius for the background (which
is low).  We created response files, and binned the spectral files
such that the bin width was at least 25\,eV (about one third of the
spectral resolution) and the number of counts was at least~25.

We first fit our 11 new observations with an absorbed blackbody
model over the 0.2--1.5\,keV range (we use \texttt{sherpa} and the
{\tt xstbabs} absorption model of \citealt{wam00}).  The best-fit
model had column density $N_{\rm H}=\expnt{(2.28\pm0.09)}{20}\,{\rm
  cm}^{-2}$, effective temperature $kT_{\rm
  BB}^{\infty}=104.0\pm0.4\,$eV, and blackbody radius $R_{\rm
  BB}^{\infty}=3.10\pm0.04 (d/{\rm500\,pc})\,$km, where the blackbody
parameters are those measured by an observer at infinity.  The results
are similar to those of \citetalias{zct+05} for the 2004 data, with
the slight offset likely resulting from changes in the EPIC
calibration and from the inclusion of the MOS data (for the
reprocessed 2004 EPIC-pn data, we find values much closer to those
given above).  The fit is reasonable, with $\chi^2=469.0$ for 431 dof.
The residuals, however, appear to show some systematic deviations,
with dips near 0.75\,keV and 0.4\,keV.  Overall, the absorbed and
unabsorbed fluxes in the 0.2--2.0\,keV band are $\expnt{2.7}{-12}$ and
$\expnt{4.0}{-12}{\rm \,erg\,s^{-1}\,cm^{-2}}$.

The reduced $\chi^2$ for the individual observations ranged from 0.9
to 1.5, with $\approx 35$ dof each, suggesting the spectra are
consistent with each other.  Indeed, fitting for both the blackbody
temperature and normalization of each observation (including that from
2004), but keeping $N_{\rm H}$ fixed, does not decrease $\chi^2$
significantly: $\Delta \chi^2=13$ for 22 fewer dof.  The changes in
temperature found are between $-0.9$\% to $+1.3$\%, comparable to the
typical 2\% uncertainties for the individual observations.  The
changes are anti-correlated with those in normalization, with $R_{\rm
  BB}^2 \propto kT_{\rm BB}^{-4}$, i.e., conserving flux.  Thus, we
see no evidence for variability, and can confidently exclude the types
of changes observed for \rxjk\ \citep{dvvmv04,vdvmv04}.

\begin{deluxetable}{cc}
\tablewidth{0pt}
\tablecaption{Measured and Derived Timing Parameters for \RXJ\label{tab:ephem}}
\tablehead{
\colhead{Quantity} & Value \\
}
\startdata
Dates (MJD) \dotfill & 54,234--54,605\\
$t_{0}$ (MJD)\dotfill        &54,383.648930(2) \\
$\nu$ (Hz) \dotfill          &0.1060644595(11) \\
$\dot \nu$ ($10^{-16}\,$\Hzsec)&$-4.6(20)$ \\
TOA rms (s) \dotfill         & 0.3 \\
$\chi^2$/dof \dotfill        & 6.0/8\\
$P$ (s)\dotfill              & 9.42822888(9)\\
$\dot P$ ($10^{-14}\,$\secsec)\dotfill   & 4.1(18)\\
$\tau_{\rm char}$ (Myr)\dotfill& 3.7\\
$B_{\rm dip}$ ($10^{13}\,$G) \dotfill   &2.0 \\
$\dot E$ ($10^{30}\,{\rm erg}\,{\rm s}^{-1}$)\dotfill&1.9\\
\enddata
\tablecomments{Quantities in parentheses are the formal 1-$\sigma$
  uncertainties on the last digit.  $\tau_{\rm char}=P/2{\dot P}$ is
  the characteristic age, assuming an initial spin period $P_0\ll P$
  and a constant magnetic field; $B_{\rm
    dip}=\expnt{3.2}{19}\sqrt{P{\dot P}}$ is the magnetic field
  inferred assuming spin-down by dipole radiation; $\dot
  E=\expnt{3.9}{46}{\dot P}/P^3$ is the spin-down luminosity.}
\end{deluxetable}

Returning to fits to the 2007--2008 data with a single model, we
examined whether the broad absorption feature at 0.75\,keV was present
in our new observations.  We modeled the absorption by a
multiplicative Gaussian.
First, constraining the absorption to be centered at 0.75\,keV with a
full width at half maximum of 0.06\,keV, as used in
\citetalias{zct+05}, we find a best-fit fractional depth of $A_{\rm
  abs}=22\%\pm3$\%, with the remaining parameters similar to the
blackbody fit
and $\chi^2=408.2$ for 430 dof.  Starting with the same values, but
now fitting for the central energy and width, we find essentially
identical results: a best-fit energy of 0.756\,keV and width of
0.073\,keV, and $\chi^2=407.0$ for 428 dof.  A reduction of 61 in
$\chi^2$ is highly significant: an F test\footnote{Using an F test for
  finding lines is incorrect when the line model is located at the
  boundary of parameter space \citep{pvdc+02}.  This, however, is not
  the case here since $A_{\rm abs}$ can have either sign.}  gives a
probability of random occurrence of $\expnt{4}{-13}$.  To verify this,
we simulated 1000 sets of 11 observations like ours using the best-fit
blackbody parameters, and fitted those to blackbody models with and
without absorption.  We found that the distribution of $A_{\rm abs}$
was centered at zero with a root-mean-square spread of 3\%, similar to
the uncertainty on $A_{\rm abs}$ in our fit to the actual data.  The
distribution of $\chi^2$ also conforms to expectations, with the
$\chi^2$ including absorption never differing from the blackbody
$\chi^2$ by more than 8.5 (compared to 61 for the real data).  Thus,
we confirm the detection of absorption at 0.75\,keV by
\citetalias{zct+05}.  Using the same continuum model but letting the
absorption depth vary for each observation, we again find no
statistically significant variability ($\Delta \chi^2=10$ for 11 fewer
dof).

If we vary the energy of the absorption line $E_{\rm abs}$ over the
range 0.3--1.0\,keV (avoiding the edges of the spectrum, where the
continuum fit and the absorption are highly covariant), we find that,
with one exception, only for $E_{\rm abs}\simeq0.75\,$keV is there
significant absorption.  Otherwise, $\Delta\chi^2<10$ and $A_{\rm
  abs}$ is around 0.  The one exception is a hint of absorption at
0.42\,keV, which gives $\Delta \chi^2=15$ and $A_{\rm
  abs}=9.7\%\pm2.3$\%.  This detection is significant at the
$\sim\!3\sigma$ level given the 15 trials that we did (none of our
1000 simulations achieved such a high $\Delta \chi^2$).  Such a line
also improves the fit for the 2004 data, although not by a
statistically significant amount ($\Delta \chi^2=4$).
Confirmation of this with the EPIC-MOS and RGS data is ongoing.

\begin{figure}
\plotone{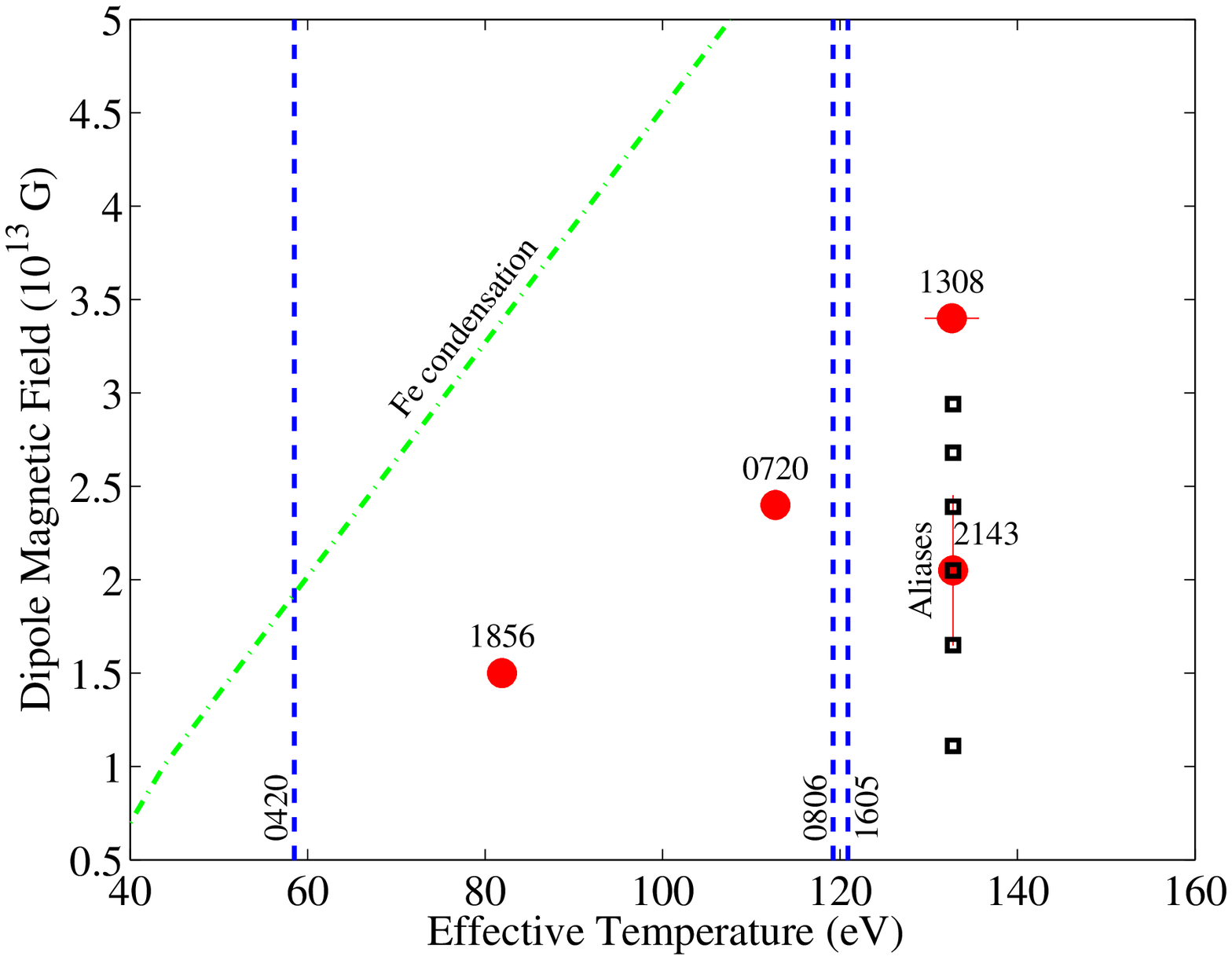}
\caption{Dipole magnetic field  vs.\
  effective temperature for the seven INS.  Temperatures are derived
  from blackbody fits including absorption features, and have been
  corrected for a nominal gravitational redshift of 0.3.  They are
  taken from \citet{bhn+03}, \citet{dvvmv04}, \citet{vkkd+04},
  \citet{shhm05}, \citetalias{zct+05} (which is consistent with the
  results from Section~\ref{sec:spec}), and \citet{hmz+04}.  The magnetic
  fields are those estimated from spin-down, and are taken from
  \citetalias{kvk05},b and \citetalias{vkk08};
  sources without spin-down solutions are shown as vertical lines.
  For \rxj, the main solution from Table~\ref{tab:ephem} is shown, as
  well as the possible aliases including the 2004 data
  (Fig.~\ref{fig:resids}).  The diagonal line is where a surface of
  iron would condense \citep{ml07b}.  Note the importance of using the
  spin-down derived magnetic fields; the inferences made in less
  secure ways can be drastically different (compare to Fig.~1 in
  \citealt{tzd04} and \citealt{plmg07}).}\label{fig:bvst}
\end{figure}

\section{Discussion and Conclusions}
A large magnetic field of $\gtrsim\!10^{14}\,$G is
inferred\footnote{This interpretation ignores  the
suppression of absorption lines due to vacuum resonance mode
conversion in such strong magnetic fields; see \citet{hl04}.} for
\rxj\ if one interprets the 0.75\,keV absorption in its spectrum as
arising from either proton cyclotron absorption ($\expnt{1.4}{14}\,$G
for a gravitational redshift $z=0.3$) or ionization of hydrogen (even
higher, as in \citetalias{zct+05}; also see \citealt{vkk07}).  This
scenario appeared to work well for \rxjk\ and \rbs, where those
transitions could match the observed absorption features for magnetic
fields of a few times $10^{13}\,$G, comparable to what was
inferred from timing (\citealt{vkk07,haberl07}, and references
therein).  It also seemed  consistent with the lack of
features in \rxjw, since that source has the weakest field
\citepalias{vkk08}.

For \rxj, though, the model breaks down: the required strong magnetic
field is inconsistent with our timing measurements ($B_{\rm
dip}=\expnt{2}{13}\,$G) at the $10^{-4}$ level (given $\Delta
\chi^2>20$ for 3 parameters).  In principle, the possible absorption
feature at 0.4\,keV may be easier to accommodate, although it still
occurs at higher energy than the features seen in \rxjk\ and \rbs\
while the magnetic field that we measure here is nominally weaker than
the fields of those sources (Fig.~\ref{fig:bvst}).  Furthermore, this
leaves the 0.75\,keV feature unexplained.  It being a ``harmonic'' of
the 0.4\,keV line seems unlikely, as it is almost twice as strong,
while one would expect harmonics to be significantly weaker
(\citealt{pp76}; G.~G.~Pavlov 2007, private communication).  A better
match may be possible with an atmosphere with helium or even heavier
elements (\citealt{pb05,hm02}; \citetalias{zct+05}; \citealt{mh07});
of course, in this case hydrogen must be absent, given the rapid
gravitational settling time \citep{ai80} and the small amount of
hydrogen required to be optically thick \citep{romani87}.

Geometry may offer a partial solution.  The magnetic field that we infer from
the spin-down rate is actually the true dipolar field strength times a
function of the angle $\alpha$ between the magnetic and rotation axes.
That function is $\sin\alpha$ in the traditional vacuum dipole model
\citep{pacini67}, which would give a large range of possible true
field strengths.  More modern analyses, however, give something more
like $\sqrt{1+\sin^2\alpha}$ \citep{spitkovsky08}, and thus a range of
only 40\%.  Therefore, it seems unlikely that this can remedy the
situation.  Modeling of the pulse profile and phase-resolved
spectroscopy has the potential to constrain the geometry
\citep[e.g.,][]{zt06}, although X-ray polarimetry may be required for
unambiguous results.

Turning to the properties of the INS as a whole, our result shows that
not only do the INS cluster at long periods, but they are close in
both axes of the $P-\dot P$ plane.  The inferred magnetic fields for
\rxjw, \rxjk, \rbs, and now \rxj\ are all within a factor of 2 of each
other.  Intriguingly, we also find a possible correlation between the
effective temperature and magnetic field (Fig.~\ref{fig:bvst}; a
similar realization was made by \citealt{plmg07} for a larger but less
uniform sample, but see also \citealt{tzd04} where the use of magnetic
fields inferred from spectroscopy led to different conclusions).
Whether or not \rxj\ fits this trend, however, will require an
unambiguous timing solution.

A possible explanation for the clustering of the magnetic field
strengths, and perhaps also the correlation with temperature, is that
the fields were originally significantly stronger
($10^{14}$--$10^{15}$~G), and decayed.  Such a scenario was originally
proposed as a way to keep the INS hotter for much longer and thus make
the natal population smaller (\citealt{hk98}; also see
\citealt{cgp00}). It is unlikely that field decay contributes much to
the current thermal state, as the cooling and kinematic ages (from
tracing the INS back to birth locations) agree well and the products
of those ages and X-ray luminosities are about 100 times the energy
currently in the dipole magnetic field, but decay may still have
influenced the current field strengths.  In particular, relatively
fast field decay naturally leads to the tightly grouped periods and
magnetic fields of the INS \citep{pg07}.  It also naturally resolves
why the spin-down ages $P/2\dot P$ (which assume constant magnetic
field) are significantly longer than the kinematic ages:
$\sim\!2\,$Myr versus $0.6\,$Myr (\citealt{kvka07,vkk07};
\citetalias{vkk08}).  Finally, field decay might also lead to a
correlation between field and temperature, since both would be a
function of age (note, however, that based on kinematic ages, \rxjw\
appears younger than \rxjk\ even though the former is colder;
\citealt{kvka07}).  If field decay occurred, the INS would be the
descendants of something like the magnetars, having had stronger
fields ($\gtrsim\!\expnt{2}{14}$~G, especially in the interiors) in
the past, but with merely above average fields now.  It could point to
an evolutionary difference between the INS and the high-magnetic-field
radio pulsars \citep[e.g.,][]{ckl+00} that inhabit the same region of
the $P$-$\dot P$ diagram.

The correlation between temperature and field strength might also be
evidence for something rather different, viz., surface physics.  For a
given magnetic field, as the surface cools, it is expected to condense
\citep{ruderman74,lai01}.  For iron, the condensation line is quite
similar to the correlation we observe \citep{tzd04,ml07b}.  From
Fig.~\ref{fig:bvst}, one sees there is about a factor~2 difference in
magnetic field, but this may just reflect the difference between the
dipole and crustal fields, or between the true temperature and that
inferred from a blackbody fit (indeed, differences in the right sense
are expected from atmosphere models; see \citealt{zp02}).  If
condensation indeed plays a role, it would help determine the surface
composition (lighter elements do not condense as easily), perhaps help
understand the peculiar spectra of the INS, and, since a condensed
surface may inhibit the formation of a vacuum gap \citep{ml07b}, help
explain the lack of radio emission from the INS
\citep[e.g.,][]{kbp+08}.
It is less clear, however, what would keep the INS on the condensation
line, or make them evolve along it.

The above possibilities may well help understand what separates the
INS from ``normal'' rotation-powered middle-aged pulsars, which have
a much wider range of magnetic field strengths in a comparable sample
\citep[e.g.,][]{kaplan08}.  Understanding how the fields of these
sources evolve and how this couples to the surface temperature is also
required to derive meaningful constraints from cooling measurements
\citep{plps04}.  To see whether the fields are indeed clustered and/or
correlated with temperature, will require measurements of other
sources, and refinement of that of \rxj.

\acknowledgements Based on observations obtained with XMM-Newton, an
ESA science mission with instruments and contributions directly funded
by ESA Member States and NASA. DLK was supported by NASA through
Hubble Fellowship grant \#01207.01-A awarded by the Space Telescope
Science Institute, which is operated by the Association of
Universities for Research in Astronomy, Inc., for NASA, under contract
NAS 5-26555.

{\it Facilities:} \facility{XMM (EPIC)}


\begin{thebibliography}{}

\bibitem[{Alcock} \& {Illarionov}(1980){Alcock} \& {Illarionov}]{ai80}
{Alcock}, C. \& {Illarionov}, A. 1980, \apj, 235, 534

\bibitem[{Burwitz} {et~al.}(2003){Burwitz}, {Haberl}, {Neuh{\" a}user},  {Predehl}, {Tr{\" u}mper}, \& {Zavlin}]{bhn+03}
{Burwitz}, V., {Haberl}, F., {Neuh{\" a}user}, R., {Predehl}, P., {Tr{\"  u}mper}, J., \& {Zavlin}, V.~E. 2003, \aap, 399, 1109

\bibitem[{Camilo} {et~al.}(2000){Camilo} {et~al.}]{ckl+00}
{Camilo}, F. {et~al.} 2000, \apj, 541, 367

\bibitem[{Colpi} {et~al.}(2000){Colpi}, {Geppert}, \& {Page}]{cgp00}
{Colpi}, M., {Geppert}, U., \& {Page}, D. 2000, \apjl, 529, L29

\bibitem[{de Vries} {et~al.}(2004){de Vries}, {Vink}, {M{\' e}ndez}, \&  {Verbunt}]{dvvmv04}
{de Vries}, C.~P., {Vink}, J., {M{\' e}ndez}, M., \& {Verbunt}, F. 2004, \aap,  415, L31

\bibitem[{Haberl}(2007){Haberl}]{haberl07}
{Haberl}, F. 2007, \apss, 308, 181

\bibitem[{Haberl} {et~al.}(2004){Haberl} {et~al.}]{hmz+04}
{Haberl}, F. {et~al.} 2004, \aap, 424, 635

\bibitem[{Hailey} \& {Mori}(2002){Hailey} \& {Mori}]{hm02}
{Hailey}, C.~J. \& {Mori}, K. 2002, \apjl, 578, L133

\bibitem[{Heyl} \& {Kulkarni}(1998){Heyl} \& {Kulkarni}]{hk98}
{Heyl}, J.~S. \& {Kulkarni}, S.~R. 1998, \apjl, 506, L61

\bibitem[{Ho} {et~al.}(2007){Ho}, {Kaplan}, {Chang}, {van Adelsberg}, \&  {Potekhin}]{hkc+07}
{Ho}, W.~C.~G., {Kaplan}, D.~L., {Chang}, P., {van Adelsberg}, M., \&  {Potekhin}, A.~Y. 2007, \mnras, 375, 821

\bibitem[{Ho} \& {Lai}(2004){Ho} \& {Lai}]{hl04}
{Ho}, W.~C.~G. \& {Lai}, D. 2004, \apj, 607, 420

\bibitem[{Jansen} {et~al.}(2001){Jansen} {et~al.}]{jla+01}
{Jansen}, F. {et~al.} 2001, \aap, 365, L1

\bibitem[{Kaplan}(2008){Kaplan}]{kaplan08}
{Kaplan}, D.~L. 2008, in American Institute of Physics Conference
Series, Vol.  983, 40 Years of Pulsars: Millisecond Pulsars, Magnetars
and More, ed.  C.~{Bassa}, Z.~{Wang}, A.~{Cumming}, \& V.~M. {Kaspi}
(New York: AIP), 331

\bibitem[{Kaplan} {et~al.}(2002){Kaplan}, {Kulkarni}, {van Kerkwijk}, \&  {Marshall}]{kkvkm02}
{Kaplan}, D.~L., {Kulkarni}, S.~R., {van Kerkwijk}, M.~H., \& {Marshall}, H.~L.  2002, \apjl, 570, L79

\bibitem[{Kaplan} \& {van Kerkwijk}(2005a){Kaplan} \& {van Kerkwijk}]{kvk05}
{Kaplan}, D.~L. \& {van Kerkwijk}, M.~H. 2005a, \apjl, 628, L45

\bibitem[{Kaplan} \& {van Kerkwijk}(2005b){Kaplan} \& {van Kerkwijk}]{kvk05b}
---. 2005b, \apjl, 635, L65

\bibitem[{Kaplan} {et~al.}(2007){Kaplan}, {van Kerkwijk}, \&  {Anderson}]{kvka07}
{Kaplan}, D.~L., {van Kerkwijk}, M.~H., \& {Anderson}, J. 2007, \apj, 660, 1428

\bibitem[{Kondratiev} {et~al.}(2008){Kondratiev}, {Burgay}, {Possenti},  {McLaughlin}, {Lorimer}, {Turolla}, {Popov}, \& {Zane}]{kbp+08}
{Kondratiev}, V.~I., {Burgay}, M., {Possenti}, A., {McLaughlin},
M.~A.,  {Lorimer}, D.~R., {Turolla}, R., {Popov}, S., \& {Zane},
S. 2008, in AIP  Conference Series, Vol. 983, 40 Years of Pulsars,
ed. C.~{Bassa}, Z.~{Wang},  A.~{Cumming}, \& V.~M. {Kaspi} (New York: AIP), 348

\bibitem[{Konenkov} \& {Popov}(1997){Konenkov} \& {Popov}]{kp97}
{Konenkov}, D.~Y. \& {Popov}, S.~B. 1997, Astron.\ Lett., 23, 498

\bibitem[{Kulkarni} \& {van~Kerkwijk}(1998){Kulkarni} \& {van~Kerkwijk}]{kvk98}
{Kulkarni}, S.~R. \& {van~Kerkwijk}, M.~H. 1998, \apjl, 507, L49

\bibitem[{Lai}(2001){Lai}]{lai01}
{Lai}, D. 2001, Rev. Mod.\ Phys., 73, 629

\bibitem[{Medin} \& {Lai}(2007){Medin} \& {Lai}]{ml07b}
{Medin}, Z. \& {Lai}, D. 2007, \mnras, 382, 1833

\bibitem[{Mori} \& {Ho}(2007){Mori} \& {Ho}]{mh07}
{Mori}, K. \& {Ho}, W.~C.~G. 2007, \mnras, 377, 905

\bibitem[{Motch} {et~al.}(2003){Motch}, {Zavlin}, \& {Haberl}]{mzh03}
{Motch}, C., {Zavlin}, V.~E., \& {Haberl}, F. 2003, \aap, 408, 323

\bibitem[{Pacini}(1967){Pacini}]{pacini67}
{Pacini}, F. 1967, \nat, 216, 567

\bibitem[{Page} {et~al.}(2004){Page}, {Lattimer}, {Prakash}, \&  {Steiner}]{plps04}
{Page}, D., {Lattimer}, J.~M., {Prakash}, M., \& {Steiner}, A.~W. 2004, \apjs,  155, 623

\bibitem[{Pavlov} \& {Bezchastnov}(2005){Pavlov} \& {Bezchastnov}]{pb05}
{Pavlov}, G.~G. \& {Bezchastnov}, V.~G. 2005, \apjl, 635, L61

\bibitem[{Pavlov} \& {Panov}(1976){Pavlov} \& {Panov}]{pp76}
{Pavlov}, G.~G. \& {Panov}, A.~N. 1976, Sov.\ Phys.\ JETP, 44, 300

\bibitem[{Pons} \& {Geppert}(2007){Pons} \& {Geppert}]{pg07}
{Pons}, J.~A. \& {Geppert}, U. 2007, \aap, 470, 303

\bibitem[{Pons} {et~al.}(2007){Pons}, {Link}, {Miralles}, \&  {Geppert}]{plmg07}
{Pons}, J.~A., {Link}, B., {Miralles}, J.~A., \& {Geppert}, U. 2007, Phys.\  Rev.\ Lett., 98, 071101

\bibitem[{Protassov} {et~al.}(2002){Protassov}, {van Dyk}, {Connors},  {Kashyap}, \& {Siemiginowska}]{pvdc+02}
{Protassov}, R., {van Dyk}, D.~A., {Connors}, A., {Kashyap}, V.~L., \&  {Siemiginowska}, A. 2002, \apj, 571, 545

\bibitem[{Rea} {et~al.}(2007){Rea} {et~al.}]{rtj+07}
{Rea}, N. {et~al.} 2007, \mnras, 379, 1484

\bibitem[{Romani}(1987){Romani}]{romani87}
{Romani}, R.~W. 1987, \apj, 313, 718

\bibitem[{Ruderman}(1974){Ruderman}]{ruderman74}
{Ruderman}, M. 1974, in IAU Symposium, Vol.~53, Physics of Dense
Matter, ed.  C.~J. {Hansen} (Dordrecht: Kluwer), 117

\bibitem[{Schwope} {et~al.}(2005){Schwope}, {Hambaryan}, {Haberl}, \&  {Motch}]{shhm05}
{Schwope}, A.~D., {Hambaryan}, V., {Haberl}, F., \& {Motch}, C. 2005, \aap,  441, 597

\bibitem[{Spitkovsky}(2008){Spitkovsky}]{spitkovsky08}
{Spitkovsky}, A. 2008, in American Institute of Physics Conference
Series, Vol.  983, 40 Years of Pulsars: Millisecond Pulsars, Magnetars
and More, ed.  C.~{Bassa}, Z.~{Wang}, A.~{Cumming}, \& V.~M. {Kaspi}
(New York: AIP), 20

\bibitem[{Turolla} {et~al.}(2004){Turolla}, {Zane}, \& {Drake}]{tzd04}
{Turolla}, R., {Zane}, S., \& {Drake}, J.~J. 2004, \apj, 603, 265

\bibitem[{van Kerkwijk} \& {Kaplan}(2007){van Kerkwijk} \& {Kaplan}]{vkk07}
{van Kerkwijk}, M.~H. \& {Kaplan}, D.~L. 2007, \apss, 308, 191

\bibitem[{van Kerkwijk} \& {Kaplan}(2008){van Kerkwijk} \& {Kaplan}]{vkk08}
---. 2008, \apjl, 673, L163

\bibitem[{van Kerkwijk} {et~al.}(2004){van Kerkwijk}, {Kaplan}, {Durant},  {Kulkarni}, \& {Paerels}]{vkkd+04}
{van Kerkwijk}, M.~H., {Kaplan}, D.~L., {Durant}, M., {Kulkarni}, S.~R., \&  {Paerels}, F. 2004, \apj, 608, 432

\bibitem[{van Kerkwijk} {et~al.}(2007){van Kerkwijk}, {Kaplan}, {Pavlov}, \&  {Mori}]{vkkpm07}
{van Kerkwijk}, M.~H., {Kaplan}, D.~L., {Pavlov}, G.~G., \& {Mori}, K. 2007,  \apjl, 659, L149

\bibitem[{Vink} {et~al.}(2004){Vink}, {de Vries}, {M{\' e}ndez}, \&  {Verbunt}]{vdvmv04}
{Vink}, J., {de Vries}, C.~P., {M{\' e}ndez}, M., \& {Verbunt}, F. 2004, \apjl,  609, L75

\bibitem[{Wang}(1997){Wang}]{w97}
{Wang}, J.~C.~L. 1997, \apjl, 486, L119

\bibitem[{Wilms} {et~al.}(2000){Wilms}, {Allen}, \& {McCray}]{wam00}
{Wilms}, J., {Allen}, A., \& {McCray}, R. 2000, \apj, 542, 914

\bibitem[{Zampieri} {et~al.}(2001){Zampieri} {et~al.}]{zct+01}
{Zampieri}, L. {et~al.} 2001, \aap, 378, L5

\bibitem[{Zane} \& {Turolla}(2006){Zane} \& {Turolla}]{zt06}
{Zane}, S. \& {Turolla}, R. 2006, \mnras, 366, 727

\bibitem[{Zane} {et~al.}(2004){Zane}, {Turolla}, \& {Drake}]{ztd04}
{Zane}, S., {Turolla}, R., \& {Drake}, J.~J. 2004, Adv.\ Space Res., 33, 531

\bibitem[{Zane} {et~al.}(2002){Zane} {et~al.}]{zhc+02}
{Zane}, S. {et~al.} 2002, \mnras, 334, 345

\bibitem[{Zane} {et~al.}(2005){Zane} {et~al.}]{zct+05}
---. 2005, \apj, 627, 397

\bibitem[{Zavlin} \& {Pavlov}(2002){Zavlin} \& {Pavlov}]{zp02}
{Zavlin}, V.~E. \& {Pavlov}, G.~G. 2002, in Neutron Stars, Pulsars, and  Supernova Remnants, ed. W.~{Becker}, H.~{Lesch}, \& J.~{Tr{\"u}mper}, 263

\end{thebibliography}


\end{document}